\documentstyle[epsf]{mn}

\newif\ifAMStwofonts
 \AMStwofontstrue

\def\dsy{\displaystyle}

\def\beq{\begin{equation}}
\def\eeq{\end{equation}}
\def\beqn{\begin{eqnarray}}
\def\eeqn{\end{eqnarray}}

\def\vc{v_\rmn{c}}
\def\vlsr{v_\rmn{lsr}}

\renewcommand{\d}{\rmn{d}}
\def\p{\upartial}

\def\kms{\mbox{$\,{\rm km}\,{\rm s}^{-1}$}}
\def\kpc{\mbox{$\,{\rm kpc}$}}

\def\refer#1#2#3{#1, #2, #3}

\def\aaa#1#2{\refer{A\&A}{#1}{#2}}

\def\aj#1#2{\refer{AJ}{#1}{#2}}

\def\araa#1#2{\refer{ARA\&A}{#1}{#2}}

\ifoldfss
  \newcommand{\rmn}[1] {{\rm #1}}

  \ifCUPmtlplainloaded \else
    \NewTextAlphabet{textbfit} {cmbxti10} {}
    \NewTextAlphabet{textbfss} {cmssbx10} {}
    \NewMathAlphabet{mathbfit} {cmbxti10} {} 
    \NewMathAlphabet{mathbfss} {cmssbx10} {} 
  \fi
  \ifAMStwofonts
    \ifCUPmtlplainloaded \else
      \NewSymbolFont{upmath} {eurm10}
      \NewSymbolFont{AMSa} {msam10}
      \NewMathSymbol{\upi}     {0}{upmath}{19}
      \NewMathSymbol{\umu}     {0}{upmath}{16}
      \NewMathSymbol{\upartial}{0}{upmath}{40}
      \NewMathSymbol{\leqslant}{3}{AMSa}{36}
      \NewMathSymbol{\geqslant}{3}{AMSa}{3E}

       \let\ge=\geqslant
    \fi
  \fi
\fi 

\ifnfssone
  \newmathalphabet{\mathit}
  \addtoversion{normal}{\mathit}{cmr}{m}{it}
  \addtoversion{bold}{\mathit}{cmr}{bx}{it}
  \newcommand{\rmn}[1] {\mathrm{#1}}

  \newmathalphabet{\mathbfit} 
  \addtoversion{normal}{\mathbfit}{cmr}{bx}{it}
  \addtoversion{bold}{\mathbfit}{cmr}{bx}{it}
  \newmathalphabet{\mathbfss} 
  \addtoversion{normal}{\mathbfss}{cmss}{bx}{n}
  \addtoversion{bold}{\mathbfss}{cmss}{bx}{n}
  \ifAMStwofonts
    \ifCUPmtlplainloaded \else
      \UseAMStwoboldmath
      \makeatletter
      \new@mathgroup\upmath@group
      \define@mathgroup\mv@normal\upmath@group{eur}{m}{n}
      \define@mathgroup\mv@bold\upmath@group{eur}{b}{n}
      \edef\UPM{\hexnumber\upmath@group}
      \new@mathgroup\amsa@group
      \define@mathgroup\mv@normal\amsa@group{msa}{m}{n}
      \define@mathgroup\mv@bold\amsa@group{msa}{m}{n}
      \edef\AMSa{\hexnumber\amsa@group}
      \makeatother
      \mathchardef\upi="0\UPM19
      \mathchardef\umu="0\UPM16
      \mathchardef\upartial="0\UPM40
      \mathchardef\leqslant="3\AMSa36
      \mathchardef\geqslant="3\AMSa3E

       \let\ge=\geqslant
    \fi
  \fi
\fi 

\ifnfsstwo
  \newcommand{\rmn}[1] {\mathrm{#1}}

  \DeclareMathAlphabet{\mathbfit}{OT1}{cmr}{bx}{it}
  \SetMathAlphabet\mathbfit{bold}{OT1}{cmr}{bx}{it}
  \DeclareMathAlphabet{\mathbfss}{OT1}{cmss}{bx}{n}
  \SetMathAlphabet\mathbfss{bold}{OT1}{cmss}{bx}{n}
  \ifAMStwofonts
    \ifCUPmtlplainloaded \else
      \DeclareSymbolFont{UPM}{U}{eur}{m}{n}
      \SetSymbolFont{UPM}{bold}{U}{eur}{b}{n}
      \DeclareSymbolFont{AMSa}{U}{msa}{m}{n}
      \DeclareMathSymbol{\upi}{0}{UPM}{"19}
      \DeclareMathSymbol{\umu}{0}{UPM}{"16}
      \DeclareMathSymbol{\upartial}{0}{UPM}{"40}
      \DeclareMathSymbol{\leqslant}{3}{AMSa}{"36}
      \DeclareMathSymbol{\geqslant}{3}{AMSa}{"3E}

       \let\ge=\geqslant
    \fi
  \fi
\fi 

\ifCUPmtlplainloaded \else
  \ifAMStwofonts \else 
    \def\upi{\pi}
    \def\umu{\mu}
    \def\upartial{\partial}
  \fi
\fi

\begin{document}

\title{The outer rotation curve of the Milky Way}

\author[J.J.\ Binney and W.\ Dehnen]
{	James Binney and Walter Dehnen	\\
Theoretical Physics, 1 Keble Road, Oxford OX1 3NP}

\pubyear{1997}

\maketitle

\begin{abstract}
A straightforward determination of the circular-speed curve $\vc(R)$ of the
Milky Way suggests that near the Sun, $\vc$ starts to rise approximately 
linearly with $R$. If this result were correct, the Galactic mass density would 
have to be independent of radius at $R\ga R_0$. We show that the apparent
linear rise in $\vc$ arises naturally if the true circular-speed curve is 
about constant or gently falling at $R_0<R\la2R_0$, but most tracers that
appear to be at $R\ga1.25R_0$ are actually concentrated into a ring of radius
$R_1\simeq1.6R_0$. 
\end{abstract}


\section{INTRODUCTION}
Determinations of the outer rotation curves of galaxies have revealed one of
the most remarkable discoveries ever made in astronomy: that the great
majority of the mass of a typical galaxy lies outside the luminous body of
the galaxy. Since we can explore our own galaxy in details that are likely
never to be accessible in external galaxies, it is important to determine
whether it conforms to this pattern. Observational data that bear on this 
question have been analyzed many times over the last decade (e.g., Fich \&
Tremaine 1991; Merrifield 1992; Brand \& Blitz 1993).

\begin{figure}
 \epsfxsize=21pc \epsfbox[30 320 615 700]{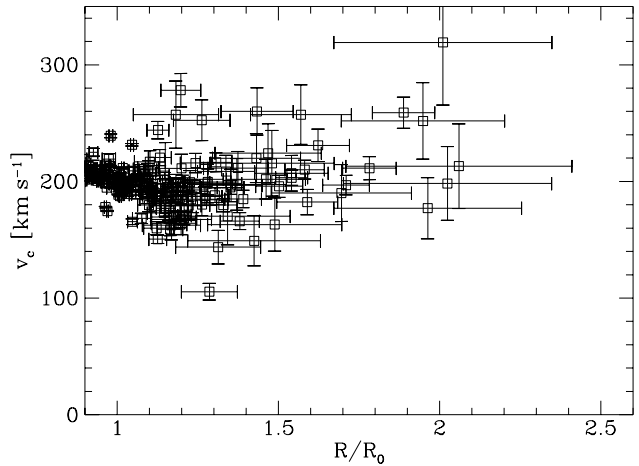}
 \caption[]{$\vc(R)$ as derived by Brand \& Blitz (1993) for the Galaxy.}
 \label{BBfig}
\end{figure}

Fig.~\ref{BBfig} shows a typical plot of $\vc(R)$ for the Galaxy -- this
example is from the study of Brand \& Blitz (1993). Two striking features of
the plot are (i) the abrupt increase in the scatter of the points and the
size of the error bars on them as one passes from $R<R_0$ to $R>R_0$, and (ii)
the clear suggestion that $\vc$ is rising linearly with $R$ at $R>R_0$ although
it is, if anything, falling with $R$ at $R\la R_0$. That this change in the
structure of the circular-speed curve occurs precisely where there is a change
in the technique used to measure it, is surely a highly suspicious circumstance.

In this note we argue that the data are naturally explained by
supposing that the observed tracers of $\vc$ are concentrated into one or
two aproximately circular rings. In this case distance errors cause the
derived circular-speed curve to rise linearly with radius irrespective of
the shape of the true form of $\vc(R)$.

In Section 2 we explain why noisy measurements of tracers that are
concentrated into a ring will simulate solid-body rotation. In Section 3 we
show that the exsting data may be successfully modelled by a ring. Section 4
sums up.  We assume throughout that $R_0=8\kpc$.

\section{WHY A RING SIMULATES SOLID-BODY ROTATION}
One estimates $\vc(R)$ at $R>R_0$ by (i) estimating the distance $d$ to
some object that is thought to be on a nearly circular orbit, and then (ii)
determining the line-of-sight velocity of this object with respect to the
local standard of rest, $\vlsr$.  The galactocentric distance $R$ of the
object can be determined from $d$\/ and the object's coordinates $(l,b)$ on the
sky, and then the circular frequency $\Omega$ at $R$ is constrained by the
formula
\beq\label{defsw}
	W(R)\equiv{\vlsr\over\sin l\cos b}=R_0\big(\Omega(R)-\Omega(R_0)\big).
\eeq
This formula makes it clear that what one actually measures is the
difference 
 \beq
	{W(R) \over R_0}=\Delta\Omega(R)\equiv\Omega(R)-\Omega(R_0)
\eeq
between the circular frequencies at the Sun and the location of the tracer. An 
error in the tracer's distance causes this difference to be associated with the
wrong radius. In particular, if all tracers were really at a single radius
$R_1$, the  scatter in the derived values of $R$ would lead one to assign the
same value of $\Delta\Omega$ to a range of values of $R$, and the outer Galaxy
would appear to be in rigid-body rotation, just as Fig.~\ref{BBfig} sugests.

\begin{figure}
 \epsfxsize=21pc \epsfbox[30 320 615 700]{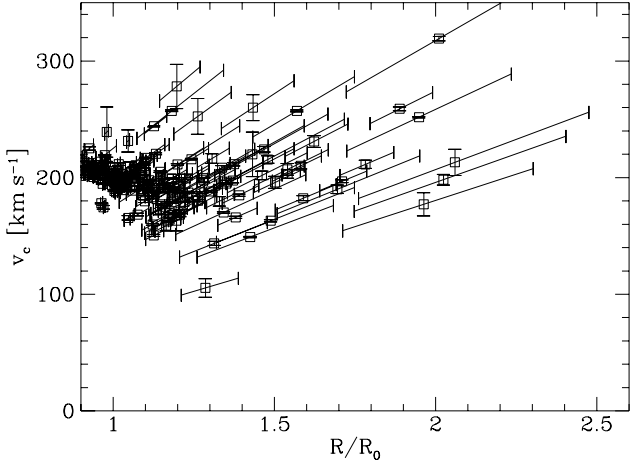}
 \caption[]{The same data as shown in Fig.~\ref{BBfig} but with the 
	    effects of errors in $W$ and $d$ shown independently.}
 \label{BB2fig}
\end{figure}

\begin{figure}
 \epsfxsize=21pc \epsfbox[30 320 615 700]{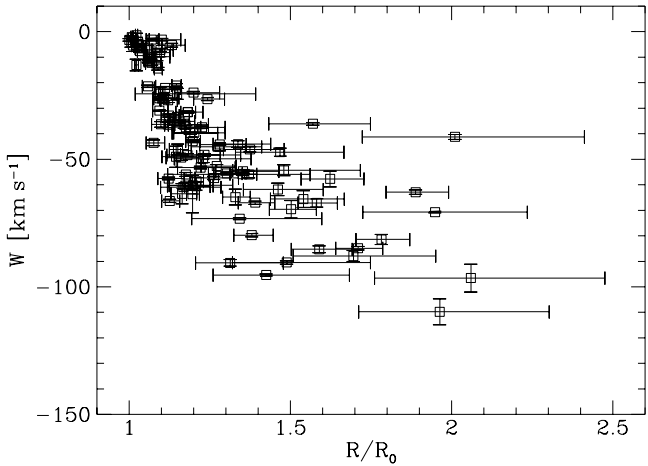}
 \caption[]{Observational values of $W(R)$ -- see equation (\ref{defsw}) 
	    -- for $R>R_0$. Tracers that lie within $15\deg$ of the anticentre
	    direction or at $|l|<90\deg$, or have estmated distances $d<1\kpc$
	    have been excluded.}
 \label{WRfig}
\end{figure}

The biasing effect of errors is obscured in Fig.~\ref{BBfig} because the
plotted errors are highly correlated. Fig.~\ref{BB2fig} shows the same data
but with the effects of errors in $W$ and in $d$\/ shown independently. 
The tendency of errors to suggest solid-body rotation is now apparent.

Fig.~\ref{WRfig} shows 99 determinations from Brand \& Blitz of the quantity
$W(R)$ that is defined by equation (\ref{defsw}). Tracers at $R<R_0$ that
have $W>0$, lie closer than $15\deg$ to the anticentre direction, or 
nearer than $1\kpc$ to the Sun have been excluded since the line-of-sight 
velocities of such objects are liable to be dominated by peculiar motion.
Fig.~\ref{WRfig} suggests that $\Omega$ drops steeply between $R_0$ and 
$\sim1.1R_0$ and is then approximately constant out to $\sim2R_0$. It
is this constancy of $\Omega$ which gives the impression of a strongly rising
circular-speed curve in Fig.~\ref{BBfig}.

\begin{figure}
  \epsfxsize=21pc \epsfbox[44 320 615 700]{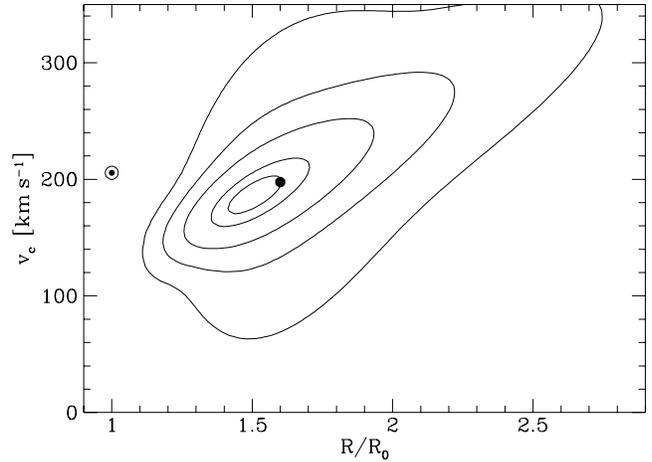}
  \caption[]{The probability of measuring $R$ and $\vc$ from objects that
	are actually distributed uniformly on a ring at $R_1=1.6R_0$.
	The contours should contain -- from inside outwards -- 
	10, 30, 68.3, 90 and 99 percent of the measurements. The observational
	errors have been assumed to be 30 per cent in distance and 12\kms\ in
	$\vlsr$ (including the instrinsic velocity dispersions of the tracers). 
	The assumed values for $\vc$ at $R_0$ and $R_1$ are indicated by 
	{$\odot$} and {\Large $\bullet$}.}
  \label{Pring}
\end{figure}

\begin{figure*}
 \hbox{ \epsfxsize=21pc \epsfbox[30 320 615 700]{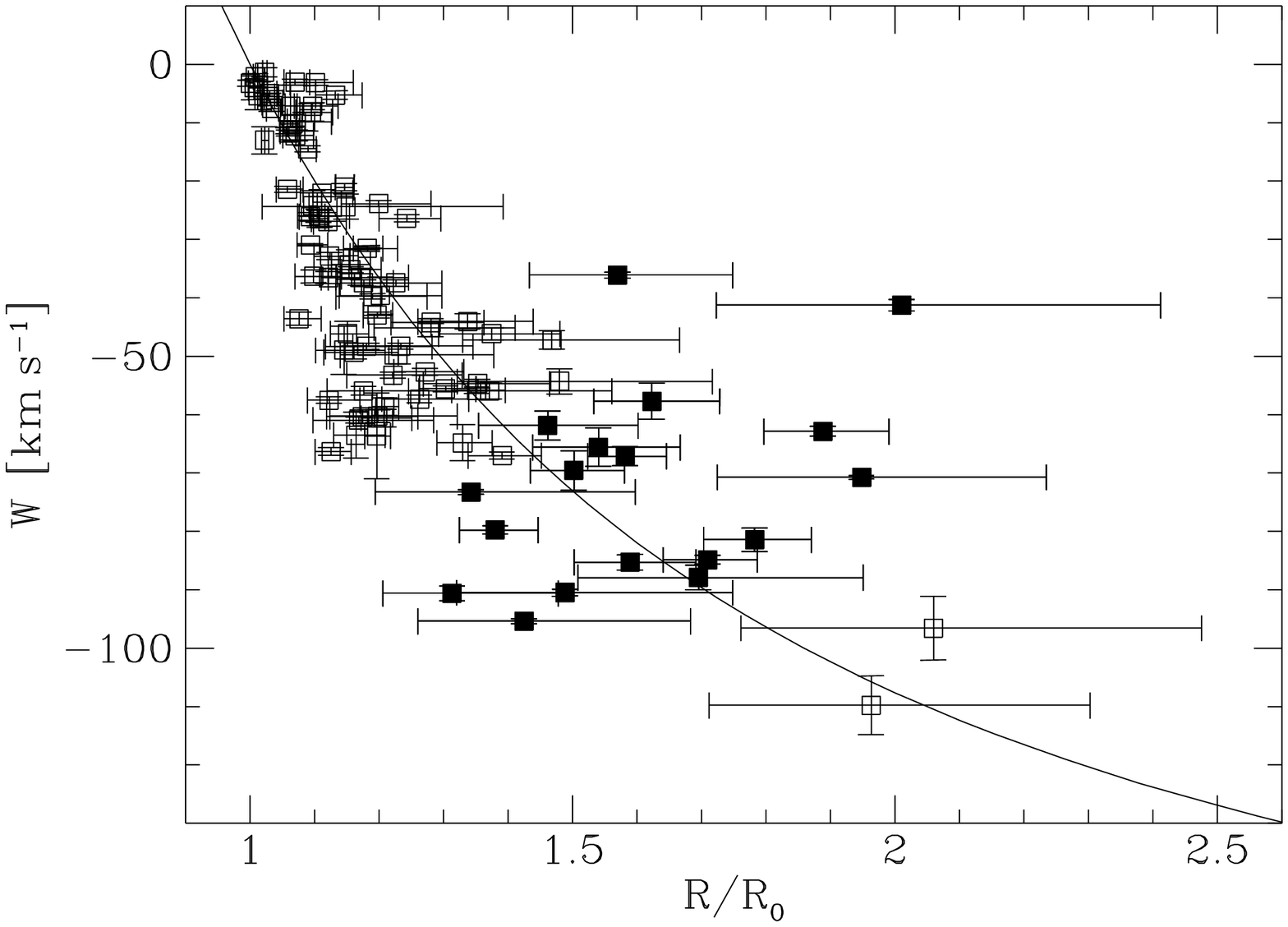}
 	\epsfxsize=21pc \epsfbox[30 320 615 700]{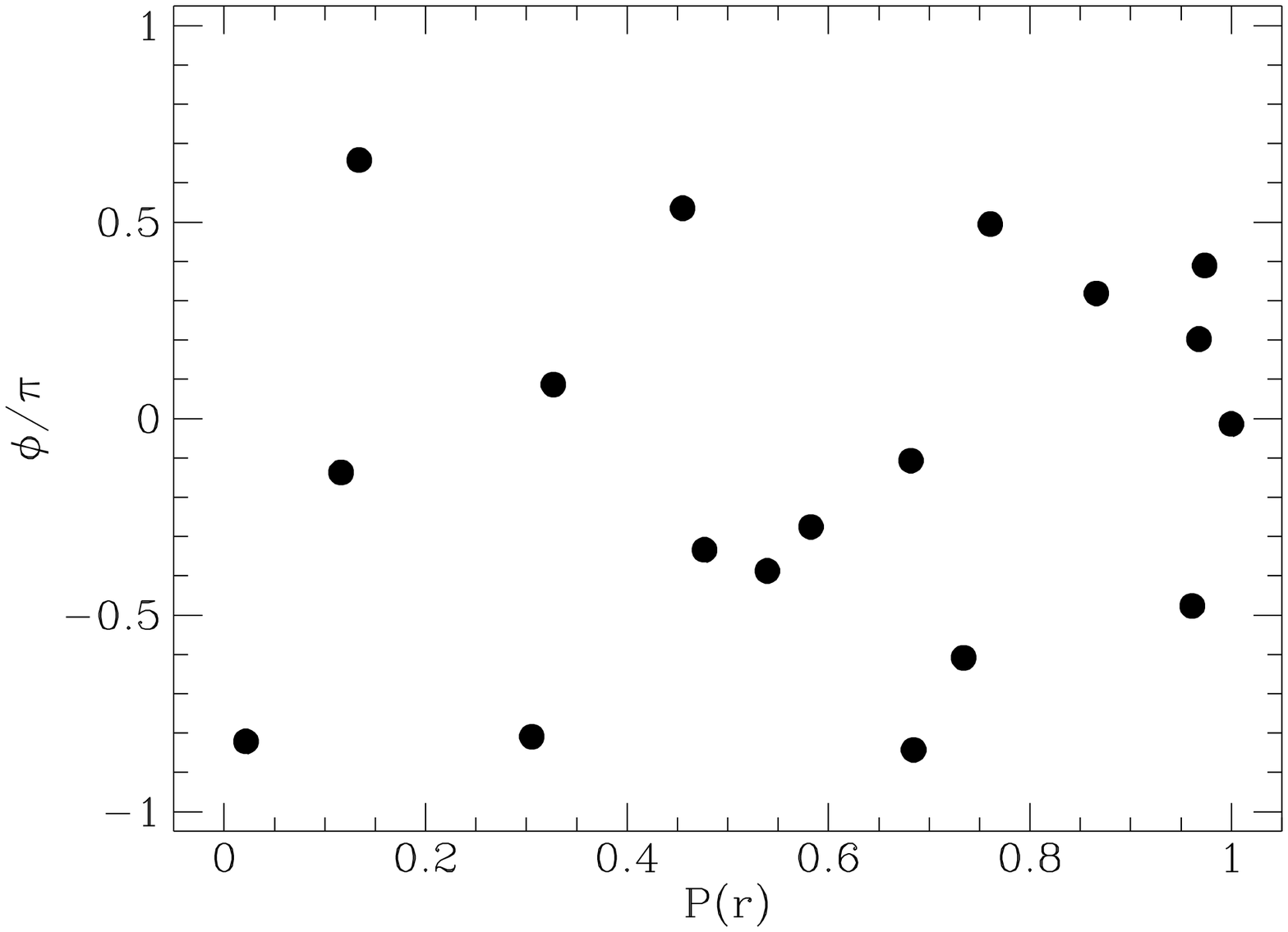} }
 \caption[]{The left panel shows the points of Fig.~\ref{WRfig} divided into
	ones that we associate with the disk near the Sun (open symbols) and
	ones that we associate with a ring of radius $1.6R_0$ (filled symbols). 
	The open symbols are compatible with $W(R)$ for Model 2 of Dehnen \&
	Binney (1997) (full curve). The right panel shows the distribution with
	$P(r)$ and $\phi$ (see text) of the points that we associate with the
	ring. This distribution should be uniform if the ring hypothesis were
	correct.}
 \label{KSfig}
\end{figure*}

\section{DO OBSERVED TRACERS LIE IN A RING?}
In view of the inherent implausibility of the result $\Omega\simeq
\hbox{constant}$ at $R>R_0$ we now investigate the possibility that
many tracers yield $\Omega\simeq\hbox{constant}$ because they lie in a ring.
Specifically, we derive for $\vc(R_0)=206\kms$ the probability of
measuring some ($R,\vc$) from objects that actually all lie on a ring at
$R_1=1.6R_0$ with $\vc(R_1)=198\kms$ (these values of $\vc$ are from
Model 2 of Dehnen \& Binney 1997). We assume that the observational errors
are normally distributed in $\ln d$ and $\vlsr$ -- that is, the probability
of measuring ($\ln d,\vlsr$) at given longitude $l$ is taken to be
\beq \label{Pringl}
	P = {\rmn{e}^{-{1\over2}(x^2+y^2)} \over
	     2\upi\;\Delta\ln d\;\Delta\vlsr }\;
	     \d \ln d_\rmn{obs}\; \d\vlsr
\eeq
with
\beq \label{xy}
	x\equiv\dsy{ {\ln d_{\rm obs}-\ln d_{\rm R}(l)\over\Delta\ln d}},\quad
	y\equiv\dsy{ {v_{\rm lsr,obs}-v_{\rm lsr,R}(l)\over\Delta \vlsr}}.
\eeq
 Here $v_\rmn{lsr,R}$ and $d_\rmn{R}$ are the line-of-sight velocity of and
distance to the ring at the longitude $l$ of a given tracer, $\Delta \ln d$
is the relative error in the distance to the tracer and $\Delta\vlsr$ is the
uncertainty in the ring's velocity at its location -- this contains
contributions from both measurement errors and the velocity dispersion of
tracers within the ring. The probability density $P(R,\vc)$ to measure
$(R,\vc)$ is then given by (\ref{Pringl}) multiplied by the Jacobian
$|\p(\ln d,\vlsr)/\p(R, \vc)|$ and convolved with an appropriate
$l$-distribution for the tracers. We obtain the latter by assuming that the
tracers are uniformly distributed around the ring in quadrants II and III,
with objects that lie within $15\deg$ of the anticentre direction excluded, as
in Fig.~\ref{WRfig}. From the outside inwards, Fig.~\ref{Pring} shows the 99,
90, 68.3, 30 and 10 percent confidence contours of $P(R,\vc)$ given that
$\Delta\ln d=0.3$ and $\Delta\vlsr=12\kms$. These contours may be
compared with the distribution of points in Fig.~\ref{BBfig}.

It is interesting to quantify further the plausibility of the ring
hypothesis.  This may be conveniently done by considering the distribution
of observed points in $x$ and $y$ (equation \ref{xy}) for a ring at
$R=1.6R_0,\vc=198\kms$. In calculating this distribution we have modified
the cited observational errors as follows: (i) we have quadratically added a
dispersion of 10\kms\ to the cited velocity errors; (ii) we have placed a
lower limit $\Delta\ln d\ge0.15$ on every distance error.  For tracers
in a ring the distribution in the $xy$ plane is given by (\ref{Pringl}),
which is constant on circles in the $xy$ plane that are centred on $(0,0)$.
Quantitatively, the probability that a tracer lies within distance $r$ of
the origin is $P(r)=1-\rmn{e}^{-r^2/2}$. Hence if the ring model is correct,
tracers that lie within the ring will be uniformly distributed in both
$P(r)$ and $\phi\equiv\arctan (x/y)$. By contrast, tracers that lie outside
the ring will clump in the $P(r)$-$\phi$ plane. Fig.~\ref{KSfig} shows a
plausible division of points between the ring and the disk near the Sun.
Specifically the left-hand panel shows the data points of Fig.~\ref{WRfig}
divided into (i) those associated with the disk near the Sun or at $R\simeq
2R_0$ (open symbols) and (ii) those associated with the ring (filled symbols).
The open symbols appear to be compatible with the curve $W(R)$ from Model 2 of
Dehnen \& Binney (1997), which is also plotted in Fig.~\ref{KSfig}. The 
right-hand panel of Fig.~\ref{KSfig} shows that the filled symbols are 
compatible with the ring hypothesis, by demonstrating that within the 
$P(r)$-$\phi$ plane these points are reasonably uniformly distributed. 
Quantitatively, one finds that the Kolmogorov-Smirnov probability that the 
points are uniformly distributed in $P(r)$ is $P_{\rm KS}=0.58$, while the 
corresponding probability for the $\phi$ distribution is $P_{\rm KS}=0.63$, 
whereas the correlation coefficient between the two is $\varrho=0.178$. 
Using Monte Carlo simulations we estimated that for a uniform distribution 
$|\varrho|\ge0.178$ with a probability of 0.48.

\section{CONCLUSIONS}
The circular-speed curve that one derives from a simple plot of
observationally estimated values of $\vc$ against $R$ is highly implausible
for two reasons. First, it abruptly reverses the sign of its derivative at
$R_0$, where there is a change in the way $\vc$ is measured. Second, at
$R>R_0$ the Galaxy appears to be rotating rigidly, which would require
the Galactic mass-density to be independent of radius at $R\ga R_0$. We
have explored the possibility that the apparent rigid rotation of the outer
Galaxy is an artifact that arises because most tracers at
$R\ga1.25R_0$ are concentrated into a ring. If there is such a
concentration of tracers, distance errors will give the erroneous result
$\vc\propto R$ because what one actually measures is $\Delta\Omega$, the
difference between the angular velocity of a tracer and that of the Sun:
everything in a ring will yield the same value of $\Delta\Omega$, and
distance errors will distribute these values in $R$.

A simple simulation of the effects of distance and velocity errors shows
that the existing data are compatible with a gently declining circular-speed
curve at $R>R_0$ provided most tracers at
$1.25R_0\la R\la2R_0$ lie in a ring of radius $\simeq1.6R_0$.

\end{document}